**Coexistence of Ferroelectric Triclinic Phases and Origin of Large Piezoelectric Responses in Highly Strained BiFeO$_3$ films**

Zuhuang Chen,[1,*] S. Prosandeev,[2] Z. L. Luo,[3] Wei Ren,[2] Yajun Qi,[1] C. W. Huang,[1] Lu You,[1] C. Gao,[4] I. A. Kornev,[5] Tom Wu,[6] Junling Wang,[1] P. Yang,[4] T. Sritharan,[1] L. Bellaiche,[2,†] and Lang Chen[1,‡]

[1] *School of Materials Science and Engineering, Nanyang Technological University, Singapore 639798, Singapore*

[2] *Institute for Nanoscience and Engineering and Physics Department, University of Arkansas, Fayetteville, Arkansas 72701, USA*

[3] *National Synchrotron Radiation Laboratory, University of Science and Technology of China, Hefei, Anhui 230029, People's Republic of China*

[4] *Singapore Synchrotron Light Source (SSLS), National University of Singapore, 5 Research Link, Singapore 117603, Singapore*

[5] *Laboratoire Structures, Proprietes et Modelisation des Solides, Ecole Centrale Paris, CNRS-UMR8580, Grande Voie des Vignes, 92295 Chatenay-Malabry Cedex, France*

[6] *Division of Physics and Applied Physics, School of Physical and Mathematical Sciences Nanyang Technological University, Singapore, 637371, Singapore*

Correspondences addressed to:

[*] m080011@e.ntu.edu.sg, [†] laurent@uark.edu , [‡] langchen@ntu.edu.sg

**Abstract**

The structural evolution of the strain-driven morphotropic phase boundary (MPB) in BiFeO$_3$ films has been investigated using synchrotron x-ray diffractometry in conjunction with scanning probe microscopy. Our results demonstrate the existence of mixed-phase regions that are mainly made up of two heavily tilted ferroelectric *triclinic* phases. Analysis of first-principles computations suggests that these two triclinic phases originate from a phase separation of a single monoclinic state accompanied by elastic matching between the phase-separated states. These first-principle calculations further reveal that the intrinsic piezoelectric response of these two low-symmetry triclinic phases is not significantly large, which thus implies that the ease of phase transition between these two energetically close triclinic phases is likely responsible for the large piezoelectric response found in the BiFeO$_3$ films near its MPB. These findings not only enrich the understandings of the lattice and domain structure of epitaxial BiFeO$_3$ films but may also shed some light on the origin of enhanced piezoelectric response near MPB.

Ferroelectric materials with morphotropic phase boundaries (MPBs) have promising potentials for industrial applications, such as transducers, actuators and sensors, because they exhibit excellent piezoelectric properties in the vicinity of MPBs [1]. Among various lead-free ferro/piezo-electric candidates, $BiFeO_3$ (BFO) is of particular interest due to its intriguing properties, such as room-temperature multiferroicity and robust piezoelectricity [2]. Recently, it has been demonstrated that epitaxial strain can be used as an alternative to chemical substitution [3] to drive so-called MPB and create large piezoelectric responses in BFO films [4-6], which revived interest in this fascinating material. Two monoclinic phases, with tetragonal-like (T-like) and rhombohedral-like (R-like) distortions, have been revealed in BFO films on substrates with lattice mismatches exceeding -4.5% [4-9]. A triclinic ferroelectric phase with large *c*/*a* ratio has been recently predicted as a low-energy metastable phase by first-principles calculations [9], but it has not been experimentally proven. Such a low-symmetry ferroelectric phase is particularly interesting for its piezoelectric response as the polarization vector is not constrained by symmetry and is thus free to rotate [10]. In addition, the detailed evolution of phase structure with film thickness, the strain-relaxation mechanism and the origin of enhanced piezoelectric properties in this strain-induced MPB system are not yet fully understood. Furthermore, it is known that the lattice structure of the films plays a crucial role in the physical properties due to the strong coupling among spin, orbital, charge, and lattice in multiferroic materials [2, 11, 12]. Therefore, it is essential to study the crystal and domain structure of highly-strained BFO films in details.

In order to gain insight into the nature of strain-induced phase transition in BFO films, we have undertaken high-resolution synchrotron x-ray diffraction (XRD) experiments, piezoelectric force microscopy (PFM) studies and first-principles calculations to investigate the crystal and domain structure of BFO films grown on LaAlO$_3$ (LAO). The complimentary experimental results show that increasing film thickness leads to a phase transformation from a pure T-like phase to a mixture of several low-symmetry phases. Besides the T-like and R-like phases, *two* other highly tilted *triclinic* ferroelectric phases have been revealed for the first time to the best of our knowledge. Furthermore, first-principles calculations we further conducted strongly suggest an interesting scenario (involving phase separation from a single monoclinic state and elastic matching) for explaining the simultaneous observation of these two triclinic phases, as well as, the enhancement of piezoelectricity in BFO films.

Epitaxial BFO films with thicknesses in the range of 10-120 nm were grown on (001) oriented LAO substrates by pulsed laser deposition [5]. Conventional $\theta-2\theta$ XRD investigations were initially done in a four-circle x-ray diffractometer (Panalytical X-pert Pro). Subsequently, high resolution XRD data were collected at Singapore Synchrotron Light Source (λ = 1.5405 Å) and Shanghai Synchrotron Radiation Facility (SSRF) (λ = 1.2398 Å). The reciprocal space maps (RSMs) were plotted in reciprocal lattice units (r.l.u.) of the LAO substrate (1 r.l.u. = 2π/3.789 Å$^{-1}$). PFM investigations were carried out on an Asylum Research MFP-3D atomic force microscope (AFM).

Figure 1(a) shows a representative AFM topography of a 10-nm-thick BFO film. Atomically flat terraces with single-unit-cell-high steps are observed, indicating the layer-by-layer growth. As shown in Fig. 1(b), only 00*l* diffraction peaks of the film and substrate were observed in the *θ*-2*θ* pattern, indicating epitaxial growth. The out-of-plane lattice parameter of BFO is calculated to be *c* ~ 4.64 Å, which demonstrates the stabilization of T-like phase by the epitaxial strain [13]. The presence of thickness fringes [Fig. 1(b)] indicates a high crystalline quality and smooth surface, consistent with the AFM data. It is known that the flat area in AFM topography correspond to the pure T-like phase and the stripe-like area correspond to the mixed phase in BFO films near its MPB [4-6, 12]. The absence of stripe-like contrast in Fig.1 (a) further suggests that the ultra-thin film consists of the T-like phase only. With increasing film thickness, the stripe-like area of the mixed phase emerges to relieve strain [4, 5]. A number of striped features are clearly observed in the topographic image of a ~80-nm-thick film in Fig. 1(c). The phase coexistence can also be identified through $\theta - 2\theta$ diffractogram, as shown in Fig. 1(d). The out-of-plane $c$ lattice parameters calculated from the position of the 00$l$ peaks are ~3.97 Å and ~4.67 Å for the R-like and T-like phases, respectively, which are in agreement with earlier studies [5, 8]. Another phase (labeled as Tri-1) with a $c$ parameter of ~4.18 Å was detected as well.

To clarify the phase structure of the film, high resolution synchrotron XRD was employed. Figure 2(a) shows a two-dimensional $HL$-plane mapping near the 002 diffraction for the 80-nm-thick film ($H$, $K$, and $L$ are reciprocal space coordinates). Diffraction peaks from the T-like ($L$ ~1.624) and R-like phases ($L$ ~1.907) were found to have the same $H$ value as that of the substrate, indicating that there is no tilt between the (001) plane of these two phases and the substrate. Besides the two peaks of the R-like and T-like phases, two additional sets of diffraction peaks were observed which included two sets of peak pairs with same $L$ value but opposite $H$ values. The first set with $L$ ~1.812 corresponds to the Tri-1 phase which has an out-of-plane lattice parameter of 4.178(1) Å and it is tilted by an angle of ±2.7° ($H$ ~ ±0.086) into the [100] direction with respect to the substrate surface normal. The second set of two peaks with $L \sim 1.619$ corresponds to another phase (labeled as Tri-2) which has an out-of-plane lattice parameter of 4.682(2) Å and a tilt angle of $\pm 1.5°$ ($H \sim \pm 0.042$) along the [100] direction. Figure 2(b) shows an asymmetric $HL$-plane mapping around the $\bar{1}03$ diffraction. It is clear that the $\bar{1}03$ diffraction peak of the T-like phase splits into three adjacent peaks as a consequence of the existence of domain variants. This demonstrates that the T-like phase is monoclinic $M_C$, which is consistent with previous studies [7-9]. This is further confirmed by in-plane domain image of the 80-nm-thick film as presented in Fig. 2(c). A stripe-like domain structure aligned along the <110> direction is clear in the flat T-like area, indicating that the polarization vector of the T-like phase lies within the (010) plane [5, 7, 8]. The lattice parameters of the T-like phase extracted from the ($\bar{1}03$) RSM are: $a_m = 3.811(1)$ Å, $b_m = 3.734(2)$ Å, $c_m = 4.670(2)$ Å and $\beta = 88.23(2)°$. Strikingly, the intensity of diffraction of the R-like phase is very weak compared with the other three phases (Tri-1, Tri-2 and T-like) and were even not detected in other studies [12], suggesting that the volume fraction of the R-like phase must be very small.

To get more detailed information on spatial arrangements of the two tilted phases, plan-view $HK$ mappings were obtained around (002) peak with $L$ values at the peaks of the two tilted phases, shown in Figs. 2(d) and (e). It is interesting to note that the (002) $HK$-plane RSMs clearly exhibits eight peaks with tilt angles along both the $H$ and $K$ directions in both figures. Note that the central strong peak in Fig. 2(e) is from the T-like phase due to the close $c$ parameters of the T-like and Tri-2 phases. The tilt angles are determined to be $\pm 2.7°$ along [100] (or [010]) direction and $\pm 0.5°$ along [010] (or [100]) direction for the Tri-1 phase, and ±1.5° along [100] (or [010]) direction and ±0.4° along [010] (or [100]) direction for the Tri-2 phase. The in-plane angles between the spots in the mapping and [100] (or [010]) direction are within 15°, which is in good agreement with previous theoretical studies [14]. The eight spots in the *HK*-plane RSMs are also consistent with the observation of eight orientations of stripe-like features on the film surface [14]. The schematic domain arrangement of the tilted phases is shown in Fig. 2(f). The tilted nature can also be verified from the TEM cross-section image (See the supplementary information). The tilt angle between the adjacent phases extracted from the TEM image is about 4°, in good agreement with the sum of tilt angles of 2.7° and 1.5° derived from the diffraction data. The presence of eight diffraction spots simultaneously in both Tri-1 and Tri-2 phases demonstrates the affinity between the two phases. Further taking into account a small fraction of the R-like phase, we suggest that the mixed-phases regions are made up not of the T-like and R-phases, as reported previously [4, 5, 8], but of an intimate mixture of the Tri-1 and Tri-2 phases. Due to low resolution along *K* direction, four variants of the Tri-1 phase are observed instead of two in the (002) *HL* mapping by conventional XRD [12]. In addition, the reason why previous studies [4, 5, 8] using conventional $\theta-2\theta$ scan did not detect the Tri-1 phase is probably due to the large tilt angle.

Combining all the diffraction data, we can unambiguously determine the lattice parameters and crystal symmetries of the two tilted phases. The basis vectors (crystallographic axes) were firstly deduced from the precisely measured coordinates for (002), ($\bar{1}$03) and (013). Both phases can be concluded as belonging to the triclinic system, from the calculations of the lengths and angles between the crystal axes. The typical lattice parameters of the two tilted phase are: $a_{Tri-1}$= 3.911(2) Å, $b_{Tri-1}$= 3.822(1) Å, $c_{Tri-1}$= 4.178(1) Å, $α_{Tri-1}$= 90.53(4)°, $β_{Tri-1}$= 90.09(7)°, and $γ_{Tri-1}$= 89.45(2)° for the Tri-1 phase; and $a_{Tri-2}$= 3.816(2) Å, $b_{Tri-2}$= 3.720(1) Å, $c_{Tri-2}$= 4.682(2) Å, $α_{Tri-2}$= 88.49(6)°, $β_{Tri-2}$= 89.78(4)°, and $γ_{Tri-2}$= 89.84(2)° for the Tri-2 phase. It is found that the unit-cell volume of the Tri-1 phase is near to that of the R-like (~62.3 Å$^3$) [8], and Tri-2 is near to that of T-like (~66.4 Å$^3$). Therefore, we deduce that the Tri-1 phase evolves from the R-like phase and is highly distorted; while Tri-2 originates from the T-like phase.

Recent theoretical studies have shown that several phases are potentially stable in highly-strained BFO films [9, 15]. In order to theoretically support the existence and coexistence of the two triclinic phases, we performed density-functional calculations (DFT) [16] using the Vienna ab initio simulation package (VASP) [17] within the local spin density approximation plus the Hubbard parameter $U$ (LSDA+$U$) with $U$ = 3.87 eV [18, 19]. We use the projected augmented wave (PAW) method and a 2×2×2 $k$-point mesh and an energy cutoff of 500 eV. We employ a 40-atoms cell, in which either a G-type or C-type antiferromagnetic (AFM) order is assumed. In order to mimic (001) epitaxial BFO films, we adopt the following lattice vectors for this 40 atom unit cell, as given in the Cartesian ($x,y,z$) setting for which the $x$, $y$ and $z$-axes are along the pseudo-cubic [100], [010] and [001] directions, respectively, by: $\vec{a}_1 = 2a(1,0,0)$, $\vec{a}_2 = 2a'(0,1,0)$, $\vec{a}_3 = 2a(\Delta_1, \Delta_2, 1+\Delta_3)$, where $a$ and $a'$ are both in-plane lattice parameters. For each considered value of these in-plane lattice constants, the $\Delta_1$, $\Delta_2$, $\Delta_3$ variables and internal atomic coordinates are relaxed to minimize the total energy, the Hellman-Feynman forces and some components of the stress tensor. Note that, in the following, we rescale our lattice parameters by a ratio of 1.0154 in order to account for the underestimation of the LSDA+U method.

Let us first concentrate on *perfect* epitaxial conditions, that is $a = a'$. Figure 3(a) shows the energy versus the in-plane lattice constant, *a*, for the *equilibrium* phases. Here, only the compressive strain region is investigated. One can see that, in agreement with previous works [20, 21], three ground-state phases are predicted to occur for these ideal epitaxial conditions: (1) a monoclinic *Cc* state with a G-type AFM ordering, for *a* ranging between 3.96 Å and 3.80 Å. This *Cc* state has both a polarization and axis about which the oxygen octahedral tilt in antiphase lying along [*uuv*] directions; (2) another monoclinic state that has also a *Cc* space group but that will be denoted as *Cc'* in the following to differentiate it from the first *Cc* phase, for *a* ranging between 3.80 Å and 3.68 Å. This *Cc'* state exhibits a large axial ratio and a large out-of-plane component polarization in addition to small in-plane components of the polarization and small oxygen octahedral tiltings. It also possesses a C-type AFM ordering, rather than a G-type as in *Cc*; and (3) a tetragonal *P4mm* phase (with also a C-type AFM ordering), which possesses a large axial ratio and a large polarization fully lying along the [001] pseudo-cubic direction, for smaller in-plane lattice constants. As one can see from Fig. 3 (a), no triclinic ground-state is found from these simulations.

On the other hand, if we allow the two in-plane lattice constants, $a$ and $a'$, to be of different magnitude and take them equal to those experimentally seen, two triclinic phases indeed emerge as ground states from these new epitaxial conditions after relaxation of the atoms and of the $\Delta_1, \Delta_2, \Delta_3$ cell variables. The energy of these two triclinic states is indicated in Fig. 3(a) for their $(a + a')/2$ average lattice constant. Table I shows the lattice vectors of these two triclinic phases, which demonstrates that they are indeed the Tri-1 and Tri-2 state experimentally observed since one triclinic state has an out-of-plane parameter around 4.1 Å (that is close to that of the corresponding parameter in the experimental Tri-1 phase) while the other triclinic phase has a predicted large out-of-plane parameter equal to 4.68 Å (exactly as in the observed Tri-2 phase). From Table I, one can also extract that the angle between the predicted $\vec{a}_3$ lattice vector and the substrate normal is predicted to be around 2° in the Tri-2 state, which is in quite close agreement with the experimental value of 1.5°. Table I also provides the Cartesian components of the polarization and of the antiferrodistortive vector whose direction represents the axis about which the oxygen octahedral tilt in antiphase fashion and whose magnitude provides the angle of such tilting [19]. We also note that the AFM ordering of the Tri-1 phase is found to be of G-type (like the *Cc* states) while it is of C-type for the Tri-2 state (like the *Cc'* phases).

Let us now try to understand why these triclinic states, rather than the *Cc* and *Cc'* phases, are experimentally observed when growing relatively thick BFO film on LAO. It is important to realize that the lattice constant of LAO is around 3.79 Å, as shown in Fig. 3(a) and yields a *Cc'* ground state. Interestingly, for this lattice constant, the energy of this particular *Cc'* phase is higher than the energy associated with the dashed line that is tangent to the energy-versus-lattice constant curves of the *Cc* and *Cc'* phases (this tangent line intercepts these curves at around 3.72 Å and 3.88 Å). Because relatively thick BFO films can relax with respect to *perfect* epitaxial conditions, it is thus energetically more favorable for (001) BFO films grown on LAO to phase separate into a *Cc* state with an in-plane lattice constant, *a* = 3.88 Å and a *Cc'* state having *a* = 3.72 Å [22]. However, these two phase-separated *Cc* and *Cc'* states have to elastically match each other since they have to coexist in the same sample (as neighboring domains). This elastic matching forces each of these states to exhibit two different in-plane parameters, *a* and *a'*, rather than a single one. As a result, this *Cc* state with $a$=3.88 Å becomes the Tri-1 state shown in Table I while the *Cc'* state with $a$=3.72 Å transforms into Tri-2 (note that this elastic matching is confirmed by realizing that the Tri-1 and Tri-2 phases have the same *a'* in-plane parameter while their average *a* parameter is also equal to this *a'* value). This explains why coexisting domains made of Tri-1 and Tri-2 phases are observed in thick-enough BFO films grown on LAO.

Let us now concentrate on the piezoelectric responses of the Tri-1 and Tri-2 phases. Table I reveals that, according to first-principles calculations, the two triclinic phases have $e_{33}$ piezoelectric coefficients that are significant but not that huge, despite the fact that their polarizations do not lie along a high-symmetry direction and is therefore free to move/rotate along any direction. The T = 0 K magnitude of these $e_{33}$ coefficients are about 2~4 C/m$^2$ which is of the same order than that in typical ferroelectrics such as tetragonal PbTiO$_3$ [23, 24]. In fact, Table I further shows that the $e_{33}$ coefficients of Tri-1 and Tri-2 are smaller than the corresponding coefficient of the *Cc* phase that has an in-plane lattice constant that is the average between the *a* and *a'* lattice constants of Tri-1 and Tri-2. On the other hand, it is likely that applying an electric field in a film in which the Tri-1 and Tri-2 phases coexist as alternating domains, will move the phase boundary between these two phases and will thus result in giant piezoelectric responses (since Tri-1 and Tri-2 have out-of-plane lattice parameters that differ by nearly 10%). Such a prediction is in line with the fact that the pure T-like and R-like phases of BFO have been found to have much smaller piezoelectricity with respect to mixed-phase samples made of alternating so-called T/R domains [6].

The ease of phase transition between the two triclinic phases can also be derived through the presence of diffuse scattering peaks connecting the diffraction peaks of these two phases, as shown in Figs. 2(a) and (b). This spread of peaks can be attributed to a transition region with lattice parameter gradient between the two tilted phases in which the polarization can rotate from the Tri-1 phase to Tri-2 phase, as shown in Fig. 3(b), which might effectively relieve epitaxial strain during phase transition. Therefore, it is not always necessary to involve dislocations to relieve misfit strain, as shown that within ten unit cells, a defect-free phase boundary has been observed with continuously changing $c$ parameters in a TEM picture [4].

In a more general point of view, it is usually difficult to clarify the mechanism of enhanced piezoelectricity at the MPB of lead oxide-based solid solutions [1], mainly due to the complex chemistry, ambiguity of structure, and disorder nature. The lattice parameters of these solid solutions [25, 26] are metrically close to cubic ($c/a \sim 1$ and $\beta \sim 90^{\circ}$) which lead to subtle differences between all possible phases, and the difficulty to distinguish them from each other, making it hard to elucidate the true crystal symmetry of MPB. On the other hand, our results suggest that highly-strained epitaxial BFO films provide an alternative perspective to verify the role of nanodomains and low-symmetry phases on the enhanced piezoelectric response for strain-induced MPB, and may lead to a better understanding of other types of MPBs. Despite the presence of nanodomains [5] and its monoclinic nature [7], pure T- or R-like phase BFO films exhibit much smaller piezoelectric properties than mixed-phase BFO films [6]. Therefore, our works suggest that the sole presence of monoclinic phase or nanodomains is not sufficient to provide large piezoelectric response. The large piezoelectric response has to be related to free energy instability [27] and to the ease of polarization rotation or phase transition under external stimuli [28]. The coexisting lowest-symmetry triclinic phases are able to bridge the T- and R- like phases and facilitate the phase transition in a stoichiometric single component compound BFO with a lead-free MPB, as shown in Fig. 3(b).

In summary, we provide direct evidences for the existence of two different triclinic phases in highly-strained multiferroic $BiFeO_3$ films through careful structural studies and detailed first-principles calculations. Our results suggest that the stripe-like mixed-phases regions are mainly made up of two highly tilted triclinic phases that originate from (1) a phase separation from a single monoclinic state and (2) elastic matching. We also propose that a large piezoelectric response should arise from the ease of field-induced phase transition between these two triclinic phases. These findings enrich the knowledge of the lattice and domain structure in strained $BiFeO_3$ films, and also shed some light on general mechanisms for enhanced electromechanical coupling near MPB, which may aid to develop new high-performance lead-free piezoelectrics.

**Acknowledgement**

L.C. acknowledges the support from Nanyang Technological University under Project No. AcRF RG 21/07 and MINDEF-NTU-JPP 10/12. L.B. acknowledges ONR Grants N00014-08-1-0915 and N00014-07-1-0825 (DURIP), DOE grant DE-SC0002220, and NSF grants DMR-0701558 and DMR-0080054 (C-SPIN). Some computations were made possible thanks to the MRI NSF grant 0722625 and to a Challenge grant from HPCMO of the U.S. Department of Defense. P.Y. would like to thank the support from SSLS via NUS Core Support C-380-003-003-001. The authors thank J. Íñiguez for useful discussions and beamline BL14B1 (SSRF) for providing the beam time under Project No. j10sr0092.

**Table I**

Physical properties of the Tri-1 and Tri-2 phases. For comparison, these properties are also given in a *Cc* phase for which the in-plane lattice constant is the average of the in-plane lattice parameters of the Tri-1 and Tri-2 phases. The polarization is estimated from the relaxed atomic displacements and calculated Born effective charges.

| Physical properties | Tri-1 | Tri-2 | *Cc* |
|---|---|---|---|
| 40-atom cell lattice vectors divided by 2 (Å) | (3.911, 0.000, 0.000) (0.000, 3.821, 0.000) (-0.014, -0.006, 4.072) | (3.721, 0.000, 0.000) (0.000, 3.821, 0.000) (0.107, 0.139, 4.680) | (3.818. 0.000. 0.000) (0.000, 3.818, 0.000) (-0.014, -0.014, 4.173) |
| Polarization, C/m$^2$ | (0.471, 0.399, 0.661) | (0.236, 0.293, 1.585) | (0.391, 0.391, 0.817) |
| Antiferrodistortive vector, Radian | (0.139, 0.125, 0.167) | (0.043, 0.046, 0.004) | (0.120, 0.120, 0.177) |
| Piezoelectric coefficient $e_{33}$ (C/m$^2$) | 3.6 | 1.9 | 4.9 |

**Figures**

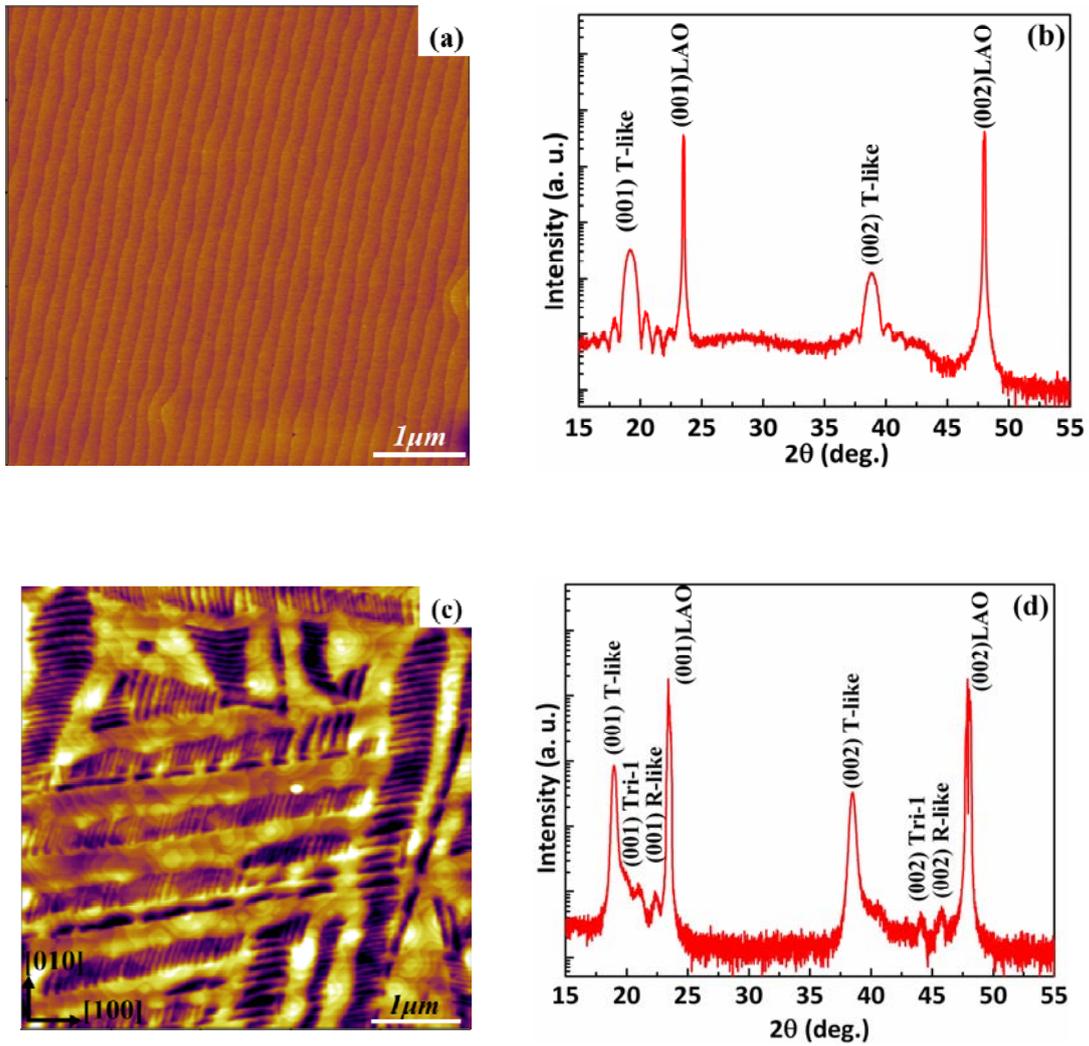

FIG. 1. (Color online) (a) AFM topography and (b) XRD θ-2θ scan of a 10-nm-thick BFO film on LAO substrate. (c) AFM topography and (d) XRD θ-2θ scan of a 80-nm-thick BFO film on LAO.

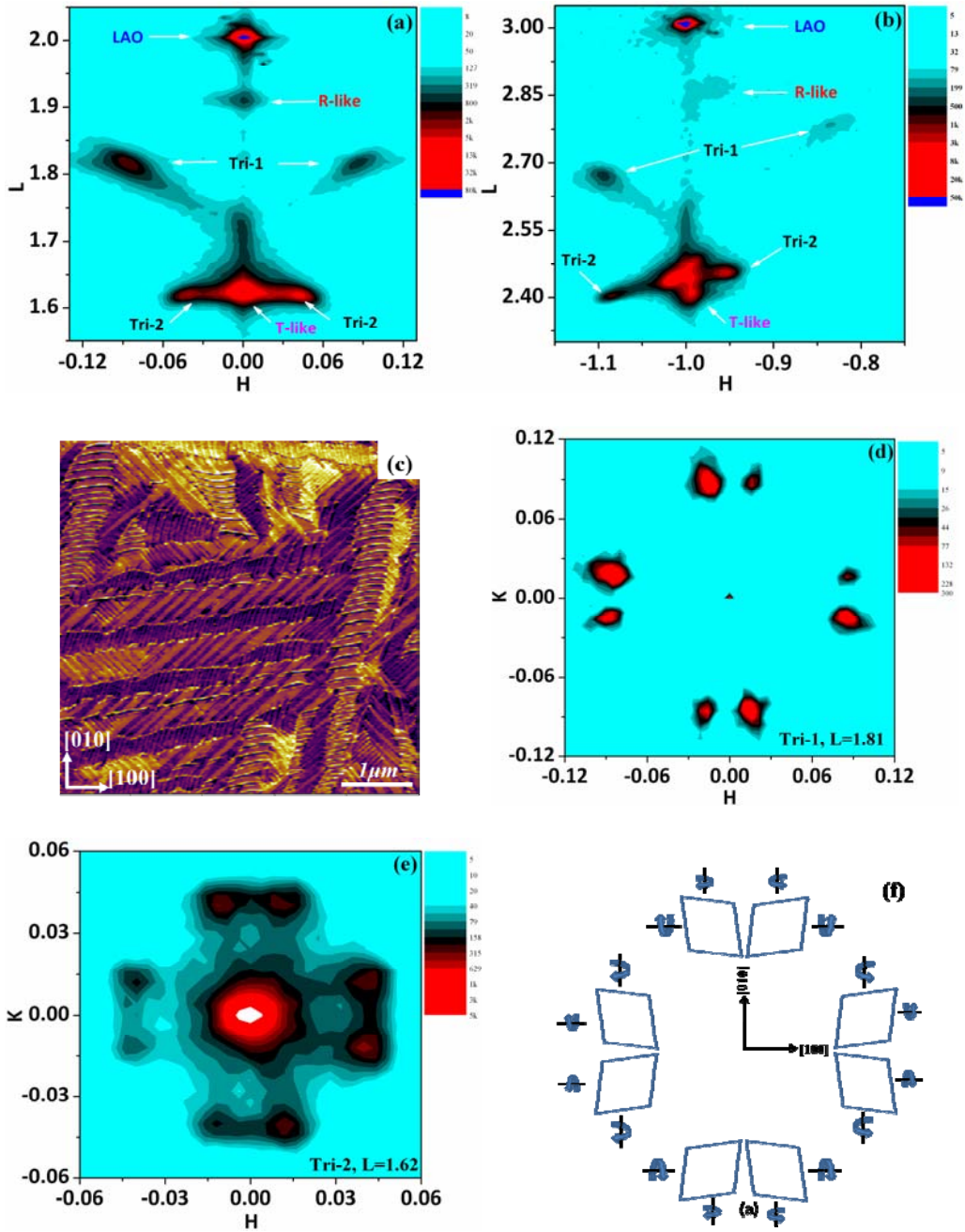

FIG. 2. (Color online) (a) (002) and (b) ($\bar{1}$03) *HL*-plane RSM of the 80-nm-thick BFO film on LAO. (c) In-plane PFM image of the 80-nm-thick BFO film. (002) *HK*-plane RSM at (d) *L*=1.81 (Tri-1 phase) and (e) *L*=1.62 (Tri-2 phase) of the film. (f) Schematic of real-space domain pattern of the tilted triclinic phases.

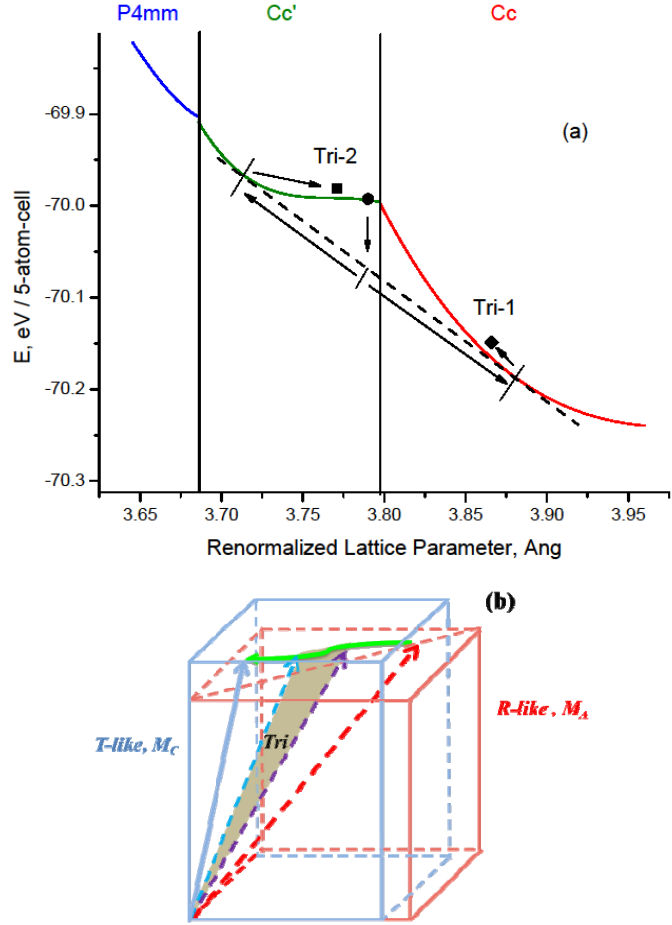

FIG. 3. (Color online) (a) Total energy versus the in-plane lattice parameter of the equilibrium phases in an epitaxial (001) BFO film, as computed from LSDA+$U$ calculations in a 40-atom cell. The red, green and blue curves represent the *Cc, Cc'* and *P*4mm phases respectively. The filled black symbols show the energies of the Tri-1 and Tri-2 phases, at their average ($a+a'$)/2 in-plane lattice parameter. The dashed line is tangent to the *Cc* and *Cc'* energy curves. The arrows schematized the proposed, phase-separated mechanism to explain the existence of the Tri-1 and Tri-2 phases. (b) Schematic of polarization rotation path between T-like $M_C$ (blue) and R-like $M_A$ (red) phases. The grey shaded region is the phase boundary between the two tilted triclinic phases.

*Supplementary Materials for the manuscript*

**Coexistence of Ferroelectric Triclinic Phases and Origin of Large Piezoelectric Responses in Highly Strained BiFeO$_3$ films**

Zuhuang Chen,[1] S. Prosandeev,[2] Z. L. Luo,[3] Wei Ren,[2] Yajun Qi,[1] C. W. Huang,[1] Lu You,[1] C. Gao,[3] I. A. Kornev,[5] Tom Wu,[6] Junling Wang,[1] P. Yang,[4] T. Sritharan,[1] L. Bellaiche,[2] and Lang Chen[1]

[1] *School of Materials Science and Engineering, Nanyang Technological University, Singapore 639798, Singapore*

[2] *Institute for Nanoscience and Engineering and Physics Department, University of Arkansas, Fayetteville, Arkansas 72701, USA*

[3] *National Synchrotron Radiation Laboratory & Department of Materials Science and Engineering, University of Science and Technology of China, Hefei, Anhui 230029, People's Republic of China*

[4] *Singapore Synchrotron Light Source (SSLS), National University of Singapore, 5 Research Link, Singapore 117603, Singapore*

[5] *Laboratoire Structures, Proprietes et Modelisation des Solides, Ecole Centrale Paris, CNRS-UMR8580, Grande Voie des Vignes, 92295 Chatenay-Malabry Cedex, France*

[6] *Division of Physics and Applied Physics, School of Physical and Mathematical Sciences Nanyang Technological University, Singapore, 637371, Singapore*

**S1 Film Synthesis**

Epitaxial BiFeO$_3$ (BFO) thin films with various thickness were grown on (001) LaAlO$_3$ (LAO) single crystal substrates (CrysTech GmbH) by pulsed laser deposition with a KrF excimer laser ($\lambda$ = 248 nm) [1]. During growth, the substrate temperature was held at 700°C in oxygen ambient of 100 mTorr. The deposition rate is around 3~5 Hz and target-substrate distance is kept at 6 cm. After deposition, the samples were slowly cooled to room temperature at a rate of 5 °C/min in 1atm of oxygen. The thicknesses of films were determined by analysis of synchrotron x-ray reflectivity (XRR) data and transmission electron microscopy.

**S2 Atomic Force Microscopy (AFM) and Piezoelectric Force Microscopy (PFM)**

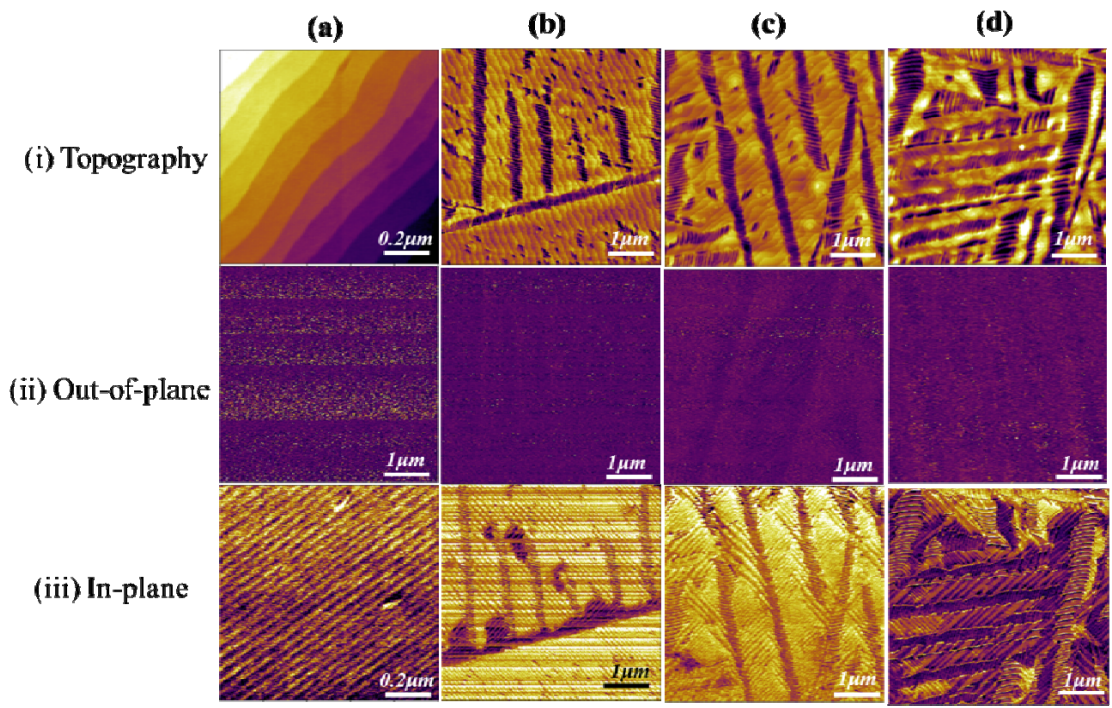

Fig. S1 (i) Topography, (ii) Out-of-plane phase, (iii) In-plane PFM images of (a) 24nm, (b) 44nm, (c) 61nm and (d) 80nm thick BFO films.

The AFM and PFM investigations were performed on an Asylum Research MFP-3D atomic force microscope using TiPt-coated Si tips (DPER18, MikroMasch). The images have been recorded with the tip cantilever pointing along <100> direction.

Figure S1 shows the topography and PFM images of BFO films with different thicknesses. As seen from topography, the areal fraction of mixed-phase area with stripe-like feature increases with increasing film thickness. Out-of-plane phase image shows uniform contrast, suggesting that all out-of-plane polarizations are pointing in one direction. The flat T-like phase regions show regular in-plane domain contrasts while in-plane PFM contrast in mixed-phase regions mimics the topography contrast. The in-plane polarization state of the T-like area should be attributed to the monoclinic distortion [1, 2].

**S3 Synchrotron X-ray Diffraction Studies**

The structural studies were completed on BFO films without bottom electrode in order to understand the direct influence of epitaxial strain on the phase evolution. The high-resolution X-ray diffraction data were obtained at beamline BL14B1 of the Shanghai Synchrotron Radiation Facility (SSRF) at a wavelength of 1.2398 Å. BL14B1 is a beamline based on bending magnet and a Si(1 1 1) double crystal monochromator was employed to monochromatize the beam. The size of the focus spot is about 0.5 mm and the end station is equipped with a Huber 5021 diffractometer which is equipped with encoders for its basic four circles ($2\theta$, $\omega$, $\chi$ and $\phi$). NaI scintillation detector was used for data collection.

To obtain the accuracy and reliability of the lattice parameters under our experimental condition at SSRF, reciprocal space vector (RSV) ($\bar{1}03$), (013) and (002) of a reference tetragonal twin-free LaSrAlO$_4$ single crystal substrate (LSAO) were measured before the real start of the experiment. (The structure scenario of films on LSAO is similar with that on LAO) The worst repeatability for their coordinates was measured to be ±2/10,000 in their *h* or *k* index. Such repeatability resulted in the errors better than 0.002 Å and 0.001 Å for their true parameters *a*=*b*= 3.756 Å and *c*= 12.636 Å, respectively; The errors for angles α, β and γ could be better than 0.02° close to 90°. This test was conducted again at the end of the experiment day, shown similar repeatability. Such accuracy throughout our experiment was reasonably good to obtain precise lattice parameters and conclude a metric crystallographic system from rest of measurements. The precision for such parameters could even be five times better.

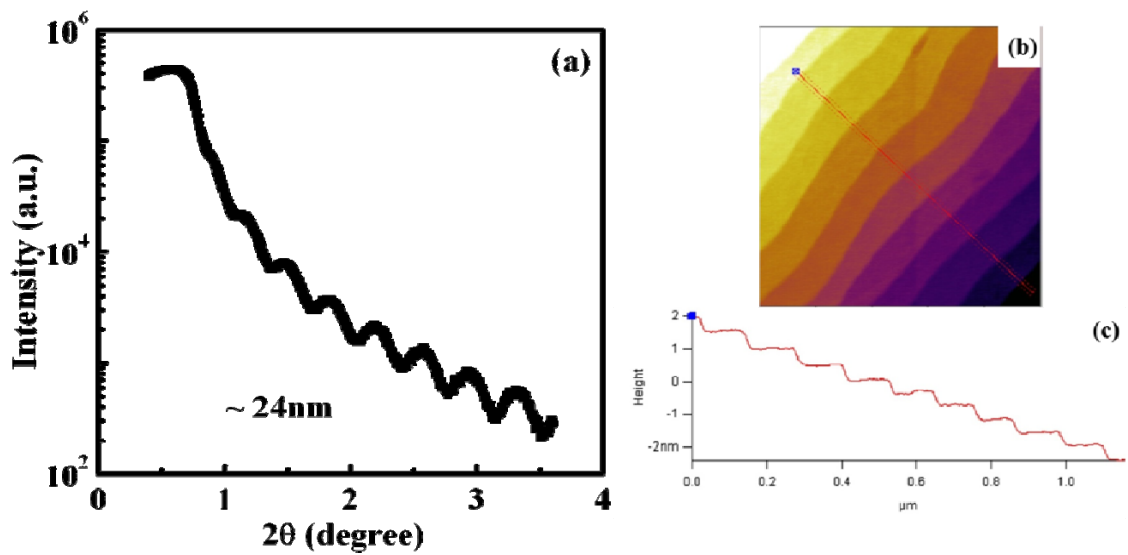

**Fig. S2 (a) A typical XRR curve of the BFO film, (b) Topography of the 24-nm-thick film, (c) Line scan profile along the line drawn in (b).**

Fig. S2a shows a typical XRR pattern. The presence of Kiessig oscillations indicates a highly ordered crystalline sample with a very smooth surface. The thickness of the film is determined to be 24 nm by the period of the intensity oscillations. The high-quality film is further confirmed by atomic force microscopy (Fig. S2b), which reveals that the surface of the film has single-unit-cell step structure (Fig. S2c) with a root-mean-square roughness ~ 0.2 nm. Figs S3 shows synchrotron x-ray reciprocal space mapping (RSM) near the (a) 002, (b) $\bar{1}03$, and (c) 114 diffractions of the 24-nm-thick film. As shown in Fig. S3a, only two diffraction peaks from LAO and T-like BFO were detected, no other tilted phases were observed indicating pure T-like phase of the film (note the presence of thickness fringe). The observation of two peaks around the $\bar{1}03$ diffraction in Fig. S3b along with the three peaks around the 114 diffraction in Fig. S3c provides clear evidence that the T-like phase adopts a monoclinic structure of $M_C$ type [2].

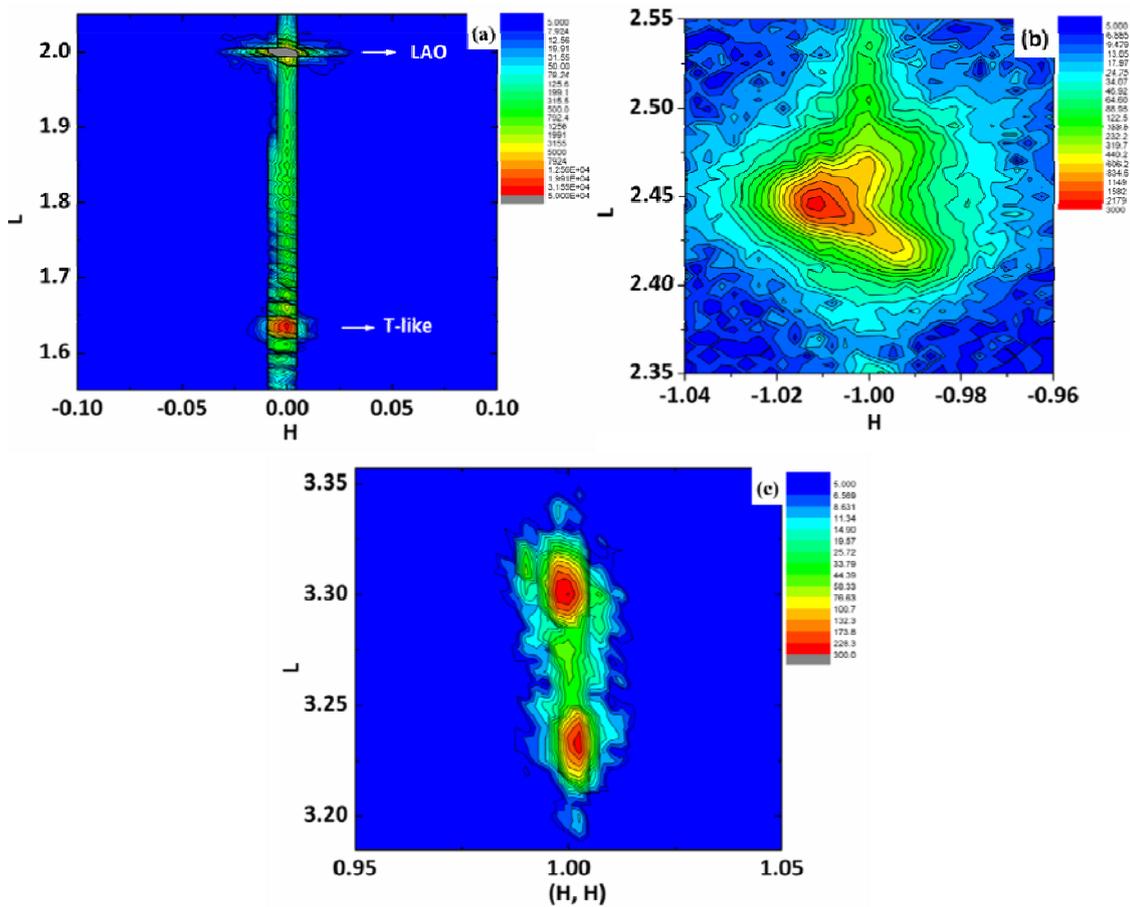

**Fig. S3 RSMs around the (a) 002, (b) $\bar{1}03$, and (c) 114 diffractions of the 24-nm-thick film.**

**S4 Transmission Electron Microscopy**

Transmission electron microscopy (TEM) studies were performed using a JEOL2100F (FEG) electron microscope operated at 200 kV. Cross-sectional TEM specimens were prepared using the standard procedure consisting of cutting, gluing, mechanical polishing, and ion milling. Fig S4 shows a typical TEM cross-section image of the 80-nm-thick BFO film on LAO. As indicated by the lines, the tilting angle between the (001) plane of the two alternating phases is about 4 degree determined from the cross-section image and SAED pattern, in agreement with the diffraction data.

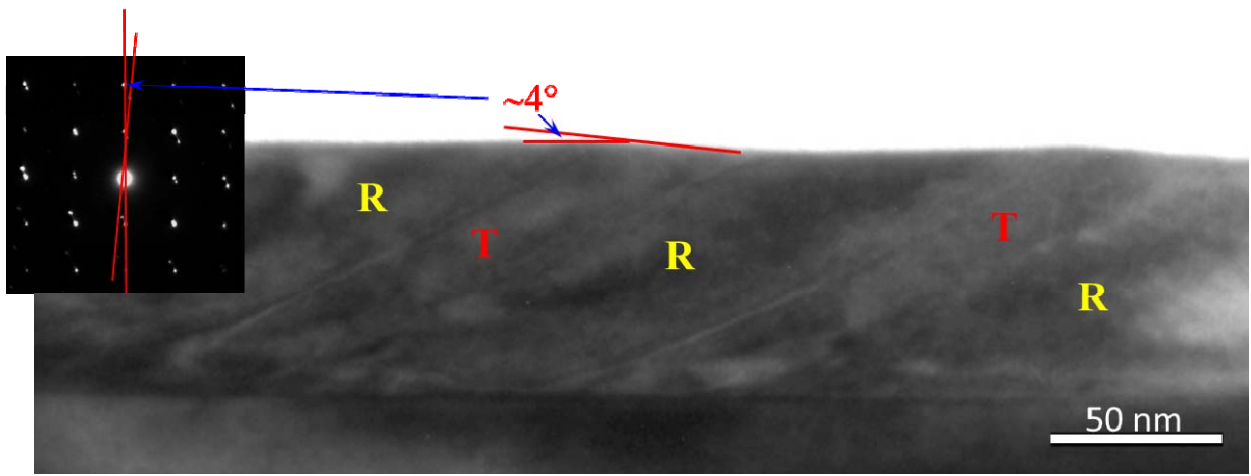

**Fig. S4 TEM cross section image of the 80-nm-thick BFO film on LAO, T and R denote to the phases with c/a ratio of 1.22 and 1.04, respectively. The inset shows the SAED pattern of the films containing the T and R phases.**